\newcommand{\version}{July 9, 2026}
\documentclass{onecolartcl}

\title{\texorpdfstring{\begin{flushright}
        {\small LA-UR-26-24767}
       \end{flushright}\vspace{2em}}{}%
       Computationally efficient method for determining limiting velocities of edge dislocations in anisotropic crystals%
       }

\author{Daniel N. Blaschke}

\date{\version}

\begin{document}

 \maketitle

 \thispagestyle{empty}
 \begin{center}
 \vspace{-0.3cm}
 Los Alamos National Laboratory, Los Alamos, NM, 87545, USA
 \\[0.5cm]
 \ttfamily{E-mail: dblaschke@lanl.gov}
 \end{center}

\begin{abstract}
The continuum-limit theory of dislocations in crystals predicts divergences in the elastic energy at crystal-geometry dependent limiting velocities $v_L$.
The $v_L$ separate subsonic, transonic, and supersonic dislocation glide regimes and are therefore important for material strength models at high strain rates.
Although it is known how to calculate these limiting velocities, there is one special case --- edge dislocations with reflection symmetry, but non-vanishing elastic constants $c'_{16}$ or $c'_{26}$ --- where previous methods have been notoriously slow.
In this letter, we address this deficiency by deriving a computationally efficient method for determining the $v_L$ of edge dislocations with reflection symmetry which is two orders of magnitude faster than the previous method.
\end{abstract}

 \vspace{1cm}
\tableofcontents
 \newpage

\section{Introduction and Background}
\label{sec:intro}

The flow stress and thereby the deformation rate of crystalline
materials, such as metals, is strongly influenced by the glide speed of dislocations \cite{Hansen:2013,Lloyd:2014,Luscher:2016,Austin:2018,Zuanetti:2021,Gurrutxaga:2020,Kim:2020,Akhondzadeh:2023,Tak:2023,Ye:2023a,Bertin:2024,Manukhina:2024}.
If dislocations reach a limiting velocity, more dislocations need to be generated in order to accommodate higher deformation rates.
Even if transonic or supersonic dislocation glide is possible, the limiting velocities constitute a hard-to-overcome barrier.
The limiting velocities $v_L$ can be determined from the continuum-limit theory of dislocations, which predicts divergences in the elastic energy at those crystal-geometry dependent $v_L$, see the review article \cite{Blaschke:2021vcrit}.
Dislocation-core regularization \cite{Markenscoff:2001,Pellegrini:2018} and lattice treatments \cite{Gurrutxaga:2025prm} have shown that these limiting velocities can be overcome in principle and molecular dynamics simulations have predicted the existence of supersonic (single) dislocations \cite{Gumbsch:1999,Olmsted:2005,Oren:2017,Dang:2022Mg,Duong:2023a,Jones:2025MSMSE} .

Recent measurements of diamond have also shown supersonic dislocations to exist \cite{Katagiri:2023}.
With the recent advances in femto-second x-ray radiography \cite{,Zelenika:2025,Borgi:2025}, this might be just the first of several future discoveries of transonic dislocations in crystals.
Aside from the questions of how dislocations can overcome (even soft) gliding velocity barriers, it is essential to know the values of these $v_L$, as the flow stress will increase non-linearly close to them \cite{Blaschke:2021impact}.
In a larger crystal plasticity finite element (CPFE) \cite{Lloyd:2014JMPS,Austin:2018,Zuanetti:2021,Ye:2023a} or discrete dislocation dynamics (DDD) simulation \cite{Cho:2015,Kim:2020,Akhondzadeh:2023,Bertin:2024}, limiting velocities will need to be calculated fast and on-the-fly as they depend on the material density and elastic constants, and are thereby temperature and pressure dependent.

Based on the so-called integral method \cite{Bacon:1980,Hirth:1982} of computing the dislocation displacement field, Barnett et al. \cite{Barnett:1973b} presented a computationally efficient method on how to determine dislocation limiting velocities in general anisotropic crystals for arbitrary slip system and dislocation character angle.
In general, there are three (slip system and dislocation character-dependent) gliding velocities where the dislocation self-energy diverges.
These limiting velocities thus separate dislocation glide into a 'sub-sonic', two 'transonic', and a 'supersonic' regime, where
the lowest of these $v_L$ separates the 'sub-sonic' from the first 'transonic' regime of dislocation glide.
However, Barnett's method does not account for special cases where the plane perpendicular to the dislocation line is a reflection plane.
In this case, the differential equations for edge and screw dislocations decouple.
Therefore the screw dislocation has only one limiting velocity and the edge dislocation has two limiting velocities.
Analytic solutions exist for the $v_L$ of screw dislocation in this case, as well as for the edge dislocation if elastic constants $c_{16}=0=c_{26}$ in the rotated coordinate system where the $\hat z$ axis is aligned with the dislocation and the $\hat y$ axis is aligned with the slip plane normal, see \cite{Bullough:1954,Teutonico:1961,Markenscoff:1984JE,Markenscoff:1984,Blaschke:2020MD} as well as the review article \cite{Blaschke:2021vcrit}.

The one case that is not well covered is an edge dislocation with reflection symmetry and non-vanishing $c_{16}$ and/or $c_{26}$.
A Python code to calculate the limiting velocity in this case was previously implemented in the present author's code PyDislocDyn \cite{pydislocdyn,Blaschke:2025joss} and the method was based on Teutonico's 1961 paper \cite{Teutonico:1961} using a combination of symbolic (sympy) and numeric methods.

In this letter, we derive a specialized version of Barnett's 1973 algorithm \cite{Barnett:1973b} adapted to the decoupled two-dimensional case of an edge dislocation with reflection symmetry.
This improved method is roughly two orders of magnitude faster to compute than the previous one that was based on Teutonico's work \cite{Teutonico:1961}.

We start by reviewing the general 3-dimensional Barnett algorithm which we subsequently simplify for the 2-dimensional case of an edge dislocation with reflection symmetry.
It must also be noted that in the case of reflection symmetry, there are subtle cancellations in the numerator of the dislocation field which depend on polar angle $\phi$ \cite{Blaschke:2020MD,Blaschke:2021vcrit}.
Therefore, the true limiting velocity can be higher than min$(v_\text{limit}(\phi))$ computed from the full 3-dimensional determinant.
In other words, the Barnett method fails in these special cases, which showcases the necessity of a 2-dimensional version specialized to the case of an edge dislocation with reflection symmetry.
Despite the simplicity of the resulting equations, this new method has not appeared in the literature to the author's best knowledge.

\section{Review of Barnett's method}
\label{sec:barnett}

The dislocation displacement field $u_i$ follows from the differential equations
\begin{align}
\partial_i \sigma_{ij}  &= \rho \ddot{u}_j
\,, &
\sigma_{ij} &= C_{ijkl} \partial_l u_{k}
\,,\label{eq:diffeqns1}
\end{align}
where $C_{ijkl}$ is the tensor of second order elastic constants (SOEC).
The established numerically efficient method of computing $u_i$ in the steady state limit, where the dislocation depends on time only via the linear combination $x-vt$, is the so-called integral method \cite{Bacon:1980,Hirth:1982}:
\begin{align}
u_j(r,\phi)&=-\frac{b_l}{2\pi}\left\{S_{jl}\ln r-S_{il}\int_0^\phi\left[(nn)^{-1}(nm)\right]_{ji}d\phi'-B_{il}\int_0^\phi(nn)^{-1}_{ji}d\phi'\right\}
\,,\nonumber\\
\mat{S}&=-\frac1{2\pi}\int_0^{2\pi}(nn)^{-1}(nm)\,d\phi  \,,\nn\\
\mat{B}&=-\frac1{2\pi}\int_0^{2\pi}\left[(mn)(nn)^{-1}(nm)-(mm)\right]d\phi
\,, \label{eq:uj-sol}
\end{align}
with the short-hand notation $(ab):=\left(a_iC_{ijkl}b_j - \rho v_ia_i\delta_{jk}v_lb_l\right)$ and we use the Einstein summation convention, i.e. indices appearing twice are summed over .
We note that all expressions in \eqref{eq:uj-sol} are $\pi$-periodic in $\phi$.
Vectors $m_i$ and $n_i$ are orthonormal to one another and lie in the plane perpendicular to the dislocation line; for details we refer to the excellent review article of Bacon, Barnett, and Scattergood \cite{Bacon:1980}.

Since \eqnref{eq:uj-sol} above depends on the inverse of matrix $(nn)$, a general mixed-character dislocation exhibits a divergence in its elastic self-energy whenever the following determinant is zero:
\begin{align}
\det \left[n_i C_{ijkl}n_l - \rho \left(v\cos\phi\right)^2\delta_{jk}\right]=0
\,,
\end{align}
where $\phi$ is the angle between velocity vector $\vec{v}$ and $\vec{n}$.
The latter vector lies in the plane perpendicular to the dislocation line and $\phi$ is the polar angle in that plane.
The determinant above can be solved in terms of $v(\phi)$ and has three $\phi$-dependent solutions in general, which are the roots of a cubic polynomial \cite{Barnett:1973b}:
\begin{align}
&x^3+bx^2+cx+d=0
\,,\nonumber\\
&b = -\tr\left(\vec{n}\cdot\mat{C}\cdot\vec{n}\right)=-\sum_{j=1}^3n_i C_{ijjl}n_l
\,,\nonumber\\
&c = \frac12\left[b^2-\tr\left(\left(\vec{n}\cdot\mat{C}\cdot\vec{n}\right)^2\right)\right] = \frac12\left[b^2-n_i C_{ijjl}n_ln_k C_{kjjr}n_r\right]
\,,\nonumber\\
&d = -\det\left(\vec{n}\cdot\mat{C}\cdot\vec{n}\right)
\,.
\end{align}
The transformation $x=y-b/3$ eliminates the quadratic term leading to the simpler equation
\begin{align}
&y^3 + 3py + 2q=0
\,,\nonumber\\
p &= \frac13\left(c-b^2/3\right)\,,&
q &= \frac{d}2-\frac{bc}{6}+\frac{b^3}{27}
\,.
\end{align}
The solutions, which are real for the elasticity problem at hand (and can be written in different ways), are
\begin{align}
y &= 2\sqrt{-p}\cos\left(\frac{\psi+2k\pi}{3}\right) & k=0,1,2
\,,\nonumber\\
\cos\psi &= \frac{-q}{\sqrt{-p^3}}
\,,\nonumber\\
\rho\left(v\cos\phi\right)^2&=x=y-b/3
\,.
\end{align}
Thus the three limiting velocities of a dislocation of mixed character angle is determined by minimizing the three branches $v(\phi)$ over the interval $[0,\pi]$ (since $\vec{n}$ is a function of $\phi$ and several terms in $\pi$-periodic \eqnref{eq:uj-sol} are integrated over $\phi$).

\section{Dislocations with reflection symmetry}

If the plane perpendicular to the dislocation line is a reflection plane, the differential equations for screw and edge dislocations decouple. Hence, one is left with a single differential equation for pure screw dislocations and two coupled differential equations for pure edge dislocations.
The equation for the screw dislocation can be solved analytically for the single limiting velocity \cite{Teutonico:1961,Blaschke:2020MD,Blaschke:2021vcrit}.
An analytic solution for the two limiting velocities of the edge dislocation exists only for a special case where one additional condition is met \cite{Teutonico:1961,Blaschke:2021vcrit}:
After rotating into a coordinate system where the $\hat z$ axis is aligned with the dislocation and the $\hat y$ axis is aligned with the slip plane normal, the elastic constants $c_{16}=0=c_{26}$ must vanish.

The more general case, where $c_{16}$, $c_{26}$ are non-zero, must be solved numerically.
A method following Teutonico's 1961 paper \cite{Teutonico:1961} was put forward in Ref. \cite{Blaschke:2021vcrit} and implemented in the Python code PyDislocDyn \cite{pydislocdyn,Blaschke:2025joss}.
However, because that implementation uses a combination of  symbolic (sympy) and (scipy) minimization routines, it is fairly slow and cannot be readily ported to C or Fortran.

On the other hand, the 3-dimensional general method of Barnett reviewed above in Section \ref{sec:barnett} fails in this high-symmetry case.
The reason is that the terms in the numerator of the dislocation field cancel out the divergence from the vanishing determinant for certain polar angles $\phi$.
In other words, it becomes unclear over what (limited) range of polar angles $\phi$ one must minimize in this case.
A prime example are the \{112\} slip planes in body-centered cubic (bcc) crystals, where reducing the minimization-interval to $[\pi/2,\pi]$ leads to the correct lowest limiting velocity for edge dislocations --- at least for the handful of cases checked by the author.

A purely numerical solution along the lines of Section \ref{sec:barnett}, but reduced to the two-dimensional problem at hand is therefore in order.
We stress that the new method we are about to derive does not require the tuning of any interval; all functions are $\pi$-periodic and therefore minimization will be strictly over $\phi\in[0,\pi]$ below.

\begin{figure}[ht]
\centering
\includegraphics[width=0.7\textwidth]{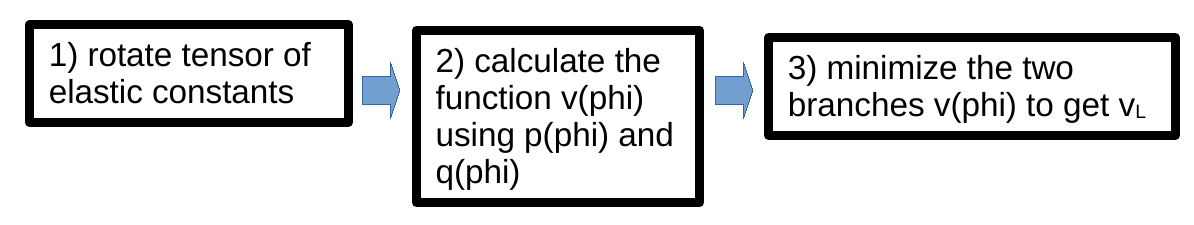}
\caption{We show a flowchart summarizing the three main steps necessary to calculate the limiting velocity of an edge dislocation with reflection symmetry using the new method presented here.
Step 1 is captured by Eq. \eqref{eq:rotateC}, step 2 two by Eq. \eqref{eq:solution} (with $p$, and $q$ given in \eqref{eq:solution_ingredients}), which is then minimized in step 3.}
\label{fig:flowchart}
\end{figure}

The first step is to rotate all quantities into a coordinate system aligned with the dislocation.
The rotation matrix we need for this purpose is determined by stacking the following three row vectors into a matrix, i.e.:
\begin{align}
	U&=\left(\begin{array}{c}
		\hat m_0^T \\ \hat n_0^T \\ \hat t^T
	\end{array}\right)
	\,, \label{eq:rotationmatrix}
\end{align}
where $\hat m_0$ is the slip direction, $\hat n_0$ is the slip plane normal, and $\hat t$ the edge dislocation line direction.
Hence, $U\cdot \hat m_0=\hat x$, $U\cdot \hat n_0=\hat y$, and $U\cdot \hat t=\hat z$.
The tensor of second order elastic constants, which is always measured in crystal coordinates, is rotated into coordinates aligned with the dislocation via
\begin{align}
	C'_{ijkl} = U_{im}U_{jn}U_{ko}U_{lp}C_{mnop}
	\,. \label{eq:rotateC}
\end{align}
The edge dislocation displacement field $u_i=U_{ij}\tilde{u}_j$ in this coordinate system has only two non-vanishing components, $u_1$ and $u_2$ (i.e. displacements only in the $x$ and $y$ directions).
Furthermore, 
\begin{align}
\vec{n}'=U_{ij}n_j = \left(\begin{array}{c}
\cos\phi\\\sin\phi\\0
\end{array}\right)
\end{align}
because of \eqref{eq:rotationmatrix}.
The determinant of the 2-dimensional matrix we must consider for the pure edge dislocation is hence
\begin{align}
\det \left[n'_i C'_{i\alpha\beta j}n'_j - \rho \left(v\cos\phi\right)^2\delta_{\alpha\beta}\right]=0
\,,
\end{align}
where $i,j\in[1,2,3]$ and free indices $\alpha$, $\beta\in[1,2]$ reflecting the 2-dimensional nature of the matrix whose determinant we presently consider.
The resulting quadratic polynomial equation in $X=\rho \left(v\cos\phi\right)^2$ is
\begin{align}
&X^2-2pX+q=0
\,,\\
p&= \frac12\left(n'_i C'_{i11 j}n'_j +n'_i C'_{i22 j}n'_j\right)
\,,\\
q &= n'_i C'_{i11 j}n'_j n'_k C'_{k22 l}n'_l - n'_i C'_{i12 j}n'_j n'_k C'_{k21 l}n'_l
\,, \label{eq:solution_ingredients}
\end{align}
where for the sake of clarity the sums over the 2D Greek indices have been written out explicitly.
The solutions are $X = p\pm\sqrt{p^2-q}$ (which are both real for the elasticity problem at hand).
Hence, the two limiting velocities for pure edge dislocations are
\begin{align}
(v_L)^2 &= \min\left[\frac{p}{\rho\cos^2\phi}\pm\sqrt{p^2-q}\right]
\,, \label{eq:solution}
\end{align}
which must be numerically minimized over $\phi\in[0,\pi]$ (since all expressions involved are $\pi$-periodic).
The lower of the two limiting velocities separates the subsonic from the transonic regime, whereas the higher limiting velocity separates the transonic from the supersonic regime, see \cite{Blaschke:2021vcrit}.
The main steps required to implement this method numerically are summarized in the flowchart of Figure \ref{fig:flowchart}.

\begin{figure}[ht]
\centering
\includegraphics[width=0.5\textwidth]{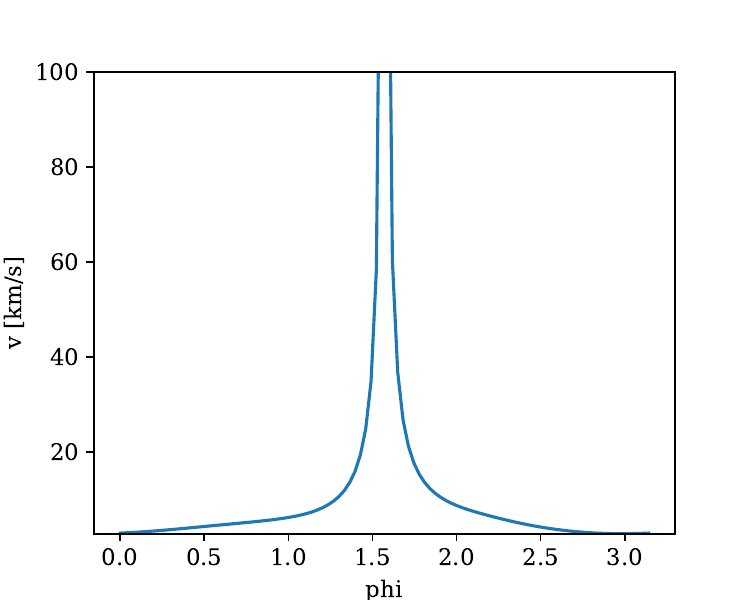}%
\includegraphics[width=0.5\textwidth]{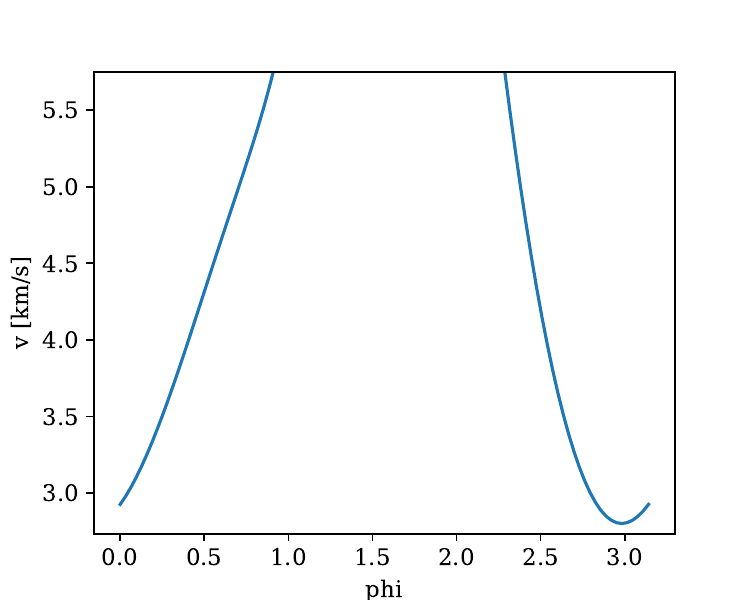}
\caption{We show the square root of the smaller branch of \eqref{eq:solution} prior to minimization at the example of a \{112\} slip plane for bcc iron.
At $\phi=\pi/2$ (where $\cos\phi=0$) the function diverges, as seen in the left panel.
In the right panel we show the same function again, but zoomed in with respect to the $y$-axis in order to see more clearly where the function's minimum lies.
In this example, the lowest $v_L=2.7999$ km/s for the edge dislocation and this minimum is found at angle $\phi\approx0.949495\pi$.}
\label{fig:vlimcurve}
\end{figure}

Figure \ref{fig:vlimcurve} shows an example of what the function we need to minimize may look like, i.e. it shows a plot of the square root of the smaller branch of \eqref{eq:solution} prior to minimization as a function of $\phi$ at the example of bcc iron for the \{112\}-slip plane.
The material density and elastic constants for Fe (and a number of other metals) are listed in Table \ref{tab:SOEC}.

\section{Results --- validating the new method}
\label{sec:results}

\begin{table}[ht]
\centering
\begin{tabular}{l|rrrrrrrr|r}
\hline
& Cr & Fe & K & Mo & Nb & Ta & V & W & Sn\\
\hline
$\rho$ [g/ccm] & 7.15 & 7.87 & 0.89 & 10.20 & 8.57 & 16.40 & 6.00 & 19.30 &  7.287\\
$c_{11}$ [GPa] & 339.80 & 226.00 & 3.70 & 463.70 & 246.50 & 260.20 & 228.70 & 522.39 & 75.29\\
$c_{12}$ [GPa] & 58.60 & 140.00 & 3.14 & 157.80 & 134.50 & 154.40 & 119.00 & 204.37 & 61.56\\
$c_{44}$ [GPa] & 99.00 & 116.00 & 1.88 & 109.20 & 28.73 & 82.55 & 43.20 & 160.58 & 21.93\\
$c_{13}$ [GPa] & - &  -&  -& - & - & - & - & - & 44.00\\
$c_{33}$ [GPa] & - &  -&  -& - & - & - & - & - & 95.52\\
$c_{66}$ [GPa] & - &  -&  -& - & - & - & - & - & 23.36\\
$A_Z$ & 0.70 & 2.70 & 6.71 & 0.71 & 0.51 & 1.56 & 0.79 & 1.01 & - \\
\hline
\end{tabular}
\caption{Material densities and elastic constants of a selection of bcc metals and body-centered tetragonal (bct) $\beta$-tin are shown.
The values are taken from Ref. \cite{CRCHandbook}, see also \cite{pydislocdyn,Blaschke:2025joss} and references therein.
The last row shows the Zener anisotropy ratio $A_Z = 2c_{44}/\left(c_{11}-c_{12}\right)$ for the cubic metals.}
\label{tab:SOEC}
\end{table}

\begin{table}[ht]
\centering
\begin{tabular}{l|l|r|r|r}
\hline
& slip system & $v_\text{lim}^\text{edge}$ [m/s], fast & $v_\text{lim}^\text{edge}$ [m/s], slow & $\left(\frac{\text{fast}}{\text{slow}}-1\right)$ \\[4pt]
\hline
Cr, (bcc) & \{112\} $<11\bar{1}>/2$ & 4092.802073 & 4092.802073 & 4e-15 \\
Fe, (bcc)  &  \{112\} $<11\bar{1}>/2$ & 2799.874438 & 2799.874438 & 4e-12 \\
K, (bcc)   & \{112\} $<11\bar{1}>/2$ & 864.896309 & 864.896309 & 8e-12 \\
Mo, (bcc)  &  \{112\} $<11\bar{1}>/2$ & 3591.693933 & 3591.693933 & 4e-15 \\
Nb, (bcc)  &  \{112\} $<11\bar{1}>/2$ & 2102.524225 & 2102.524225 & 4e-14 \\
Ta, (bcc)   & \{112\} $<11\bar{1}>/2$ & 1932.919026 & 1932.919026 & 1e-12 \\
V, (bcc)   & \{112\} $<11\bar{1}>/2$ & 2883.439916 & 2883.445660 & -2e-06 \\
W, (bcc)   & \{112\} $<11\bar{1}>/2$ & 2875.030768 & 2875.030767 & 3e-11 \\
Sn, (bct) &  \{011\} $<01\bar{1}>$ & 1681.278992 & 1681.278992 & 1e-13 \\
\hline
\end{tabular}
\caption{We compare limiting velocities in units of m/s calculated with both methods (i.e. the slow one based on Teutonico, and the fast one presented here).
The last column shows the difference between the two methods, normalized by the 'slow' one.
Small differences are expected as both algorithms rely on numeric minimization.}
\label{tab:vlim}
\end{table}

The new method has been implemented in PyDislocdyn 1.3.5 \cite{pydislocdyn,Blaschke:2025joss} and has been verified to yield comparable results to the slower previous method of Refs. \cite{Teutonico:1961,Blaschke:2021vcrit}.
As an additional consistency check, one can calculate the line tension \cite{Blaschke:2017lten} and/or drag coefficient from phonon wind \cite{Alshits:1992,Blaschke:2018anis} close to the limiting velocity to see that indeed they quickly increase in value as the correct limiting velocity is approached.
Furthermore, the new method works equally well for the special case of $c_{16}=0=c_{26}$ or even the isotropic limit (where analytic expressions exist).
In fact, PyDislocDyn uses this numerical method if $c_{16}=0=c_{26}$ and the user requested both limiting velocities (rather than just the lowest one) for an edge dislocation with reflection symmetry.

We proceed to look at a few examples with elastic constants and material densities listed in Table \ref{tab:SOEC}.
Review article \cite{Blaschke:2021vcrit} tells us that the \{112\} slip planes in bcc crystals fulfill the reflection symmetry requirement with non-vanishing (rotated) elastic constants $c_{16}$, $c_{26}$.
Table \ref{tab:vlim} compares results for the lowest limiting velocity of edge dislocations in a selection of bcc crystals using both the new (fast) method presented here, as well as the much slower one that was based on Teutonico's paper.

Tetragonal crystals also exhibit edge dislocations with reflection symmetry in many of their slip systems, most of which have vanishing elastic constants $c_{16}$, $c_{26}$.
In the case of $\beta$-tin, we have identified one slip system in the list presented in Ref. \cite{Cai:2020}, that indeed features a reflection symmetry for its edge dislocation with non-vanishing $c_{16}$ and $c_{26}$.
This is the one we have included in the last row of our examples in Table \ref{tab:vlim}.
In all cases, the results of both methods are comparable, though very small differences are expected as both algorithms rely on numeric minimization.
PyDislocDyn uses scipy's default options when employing the minimization function \verb|scipy.optimize.direct()| which is an implementation of the  DIviding RECTangles algorithm of Ref. \cite{Jones1993}.

For all metals in Table \ref{tab:vlim}, the run-time of the Python-implementation of PyDislocDyn on an M3 CPU was measured to be between 120--150 ms for the 'slow' method and 0.5--0.6 ms for the 'fast' method.
Furthermore, the latter can be implemented in a compiled language, such as Fortran or C, whereas the 'slow' method requires symbolic manipulations with sympy.

Having established that the 'fast' and 'slow' methods give comparable results (see Table \ref{tab:vlim}), it remains to show an example where the 3D Barnett method of Section \ref{sec:barnett} fails.
Choosing once more iron as our example (using the values listed in Table \ref{tab:SOEC}), the new method tells us that the two limiting velocities of the edge dislocation in the \{112\} slip plane are 2799.87 m/s and 6409.70 m/s.
The 3D Barnett method of Section \ref{sec:barnett}, on the other hand yields three values, namely 2629.48 m/s, 2925.01 m/s, and 6409.70 m/s (i.e. only the highest one is correct).

Computing the drag coefficient from phonon wind using PyDislocDyn at three different velocities (one low and the other two slightly below the predicted lowest limiting velocities of the Barnett and the 'new' methods), we find for drag coefficient $B$ of an edge dislocation in a \{112\} slip plane of iron:
\begin{align*}
B(v=300\text{m/s})&=0.0215 \text{ mPa\,s}, 
\nonumber\\
B(v=2629.48\text{m/s})&=0.1221 \text{ mPa\,s},
\nonumber\\
B(v=2799.8\text{m/s})&=144.99 \text{ mPa\,s}\,, 
\end{align*}
confirming that our new method (and not Barnett's) yields the correct result in this case since the drag coefficient diverges only at the correct limiting velocity \cite{Blaschke:2018anis}, 2799.87 m/s in this example.

\section{Conclusion}

We have derived a computationally efficient fast method of determining the limiting velocities of gliding edge dislocations with reflection symmetry, but non-vanishing elastic constants $c'_{16}$ or $c'_{26}$ --- the one special case where an efficient method was previously lacking in the literature.

A reference implementation of how to compute limiting velocities of gliding dislocations for all cases is given within the open source code PyDislocDyn \cite{pydislocdyn,Blaschke:2025joss} developed by the present author.
The method presented here for pure edge dislocations in anisotropic crystals where the plane perpendicular to the dislocation is a reflection plane, constitutes the last ``missing piece'' needed to calculate all dislocation limiting velocities for all slip systems and dislocation characters numerically efficient and without the need for any symbolic (sympy) routines.
It is therefore now possible to implement on-the-fly calculations of dislocation limiting velocities into larger simulations codes, such as crystal plasticity finite element (CPFE) or discrete dislocation dynamics (DDD) simulations.
An implementation in Fortran (in addition to Python) has therefore been included in the current development version of PyDislocDyn, i.e.
a Fortran library (or frontend) can be compiled and linked to (called from) other codes where speed is essential.

\subsection*{Acknowledgements}
\noindent
The author thanks the anonymous reviewers for valuable feedback and is grateful for the support of the Materials project within the Advanced Simulation and Computing, Physics and Engineering Models Program of the U.S. Department of Energy under contract 89233218CNA000001.

\bibliographystyle{utphys-custom}
\bibliography{dislocations}

@incollection{Alshits:1992,
author = "Alshits, Vladimir I.",
title = "The Phonon-Dislocation Interaction and its Role in Dislocation Dragging and Thermal Resistivity",
editor = "Vladimir L. Indenbom and Jens Lothe",
booktitle = "Elastic Strain Fields and Dislocation Mobility",
publisher = "Elsevier",
year = "1992",
volume = "31",
pages = "625-697",
series = "Modern Problems in Condensed Matter Sciences",
doi = "10.1016/B978-0-444-88773-3.50018-2",
}

@article{Austin:2018,
author = {Ryan A. Austin},
title = {Elastic precursor wave decay in shock-compressed aluminum over a wide range of temperature},
journal = {J. Appl. Phys.},
volume = {123},
number = {3},
pages = {035103},
year = {2018},
doi = {10.1063/1.5008280},
}

@article{Bacon:1980,
author = "D. J. Bacon and D. M. Barnett and R. O. Scattergood",
title = "Anisotropic continuum theory of lattice defects",
journal = "Prog. Mater. Sci.",
volume = "23",
pages = "51-262",
year = "1980",
doi = "10.1016/0079-6425(80)90007-9",
}

@article{Barnett:1973b,
author = {D M Barnett and J Lothe and K Nishioka and R J Asaro},
title = {Elastic surface waves in anisotropic crystals: a simplified method for calculating {Rayleigh} velocities using dislocation theory},
journal = {J. Phys. F: Met. Phys.},
doi = {10.1088/0305-4608/3/6/001},
year = {1973},
volume = {3},
number = {6},
pages = {1083-1096},
}

@article{Bullough:1954,
	author = {R. Bullough and B. A. Bilby},
	title = {Uniformly Moving Dislocations in Anisotropic Media},
	journal = {Proc. Phys. Soc.},
	doi = {10.1088/0370-1301/67/8/303},
	year = {1954},
	volume = {B67},
	number = {8},
	pages = {615-624},
}

@article{Cai:2020,
	author = {Xiaorong Cai and Carol A. Handwerker and John E. Blendell and Marisol Koslowski},
	title = {Shallow grain formation in {Sn} thin films},
	journal = {Acta Mater.},
	volume = {192},
	pages = {1-10},
	year = {2020},
	issn = {1359-6454},
	doi = {10.1016/j.actamat.2020.03.014},
}

@article{Cho:2015,
author="Cho, Jaehyun and Junge, Till and Molinari, Jean-Fran{\c{c}}ois and Anciaux, Guillaume",
title="{Toward a 3D coupled atomistic and discrete dislocation dynamics simulation: dislocation core structures and Peierls stresses with several character angles in FCC aluminum}",
journal="AMSES",
year="2015",
volume="2",
number="1",
pages="12",
doi="10.1186/s40323-015-0028-6",
}

@article{Duong:2023a,
	author = {Ta Duong and Michael J. Demkowicz},
	title = {Resonance with Surface Waves Induces Forbidden Velocity Bands in Dislocation Glide},
	year = {2023},
	journal = {J. Mech. Phys. Solids},
	volume = {180},
	pages = {105422},
	doi = {10.1016/j.jmps.2023.105422},
	OPTjournal  = {available at SSRN: 4373001},
	OPTdoi = {10.2139/ssrn.4373001},
}

@article{Gurrutxaga:2020,
author = {Be{\~{n}}at Gurrutxaga-Lerma and Jonas Verschueren and Adrian P. Sutton and Daniele Dini},
title = {The mechanics and physics of high-speed dislocations: a critical review},
journal = {Int. Mater. Rev.},
volume = {66},
number = {4},
pages = {215-255},
year  = {2021},
publisher = {Taylor \& Francis},
doi = {10.1080/09506608.2020.1749781},
}

@article{Gurrutxaga:2025prm,
	title = {Quantum statistical theory of dislocation mobility in discrete lattices},
	author = {Be{\~{n}}at Gurrutxaga-Lerma},
	journal = {Phys. Rev. Mater.},
	volume = {9},
	issue = {12},
	pages = {123605},
	numpages = {40},
	year = {2025},
	publisher = {American Physical Society},
	doi = {10.1103/zzct-418n},
}

@article{Hansen:2013,
author = "B. L. Hansen and I. J. Beyerlein and C. A. Bronkhorst and E. K. Cerreta and D. Dennis-Koller",
title = "A dislocation-based multi-rate single crystal plasticity model",
journal = "Int. J. Plast.",
volume = "44",
pages = "129-146",
year = "2013",
issn = "0749-6419",
doi = "10.1016/j.ijplas.2012.12.006",
}

@article{Kim:2020,
	author = "Soon Kim and Hokun Kim and Keonwook Kang and Sung Youb Kim",
	title = "Relativistic effect inducing drag on fast-moving dislocation in discrete system",
	journal = "Int. J. Plast.",
	volume = "126",
	pages = "102629",
	year = "2020",
	doi = "10.1016/j.ijplas.2019.11.008",
}

@article{Lloyd:2014JMPS,
title = "Plane wave simulation of elastic-viscoplastic single crystals",
journal = "J. Mech. Phys. Solids",
volume = "69",
pages = "14-32",
year = "2014",
doi = "10.1016/j.jmps.2014.04.009",
author = "J. T. Lloyd and J. D. Clayton and R. A. Austin and D. L. McDowell",
}

@article{Lloyd:2014,
author = "J. T. Lloyd and J. D. Clayton and R. Becker and D. L. McDowell",
title = "Simulation of shock wave propagation in single crystal and polycrystalline aluminum",
journal = "Int. J. Plast.",
volume = "60",
pages = "118-144",
year = "2014",
doi = "10.1016/j.ijplas.2014.04.012",
}

@article{Luscher:2016,
author = "D. J. Luscher and J. R. Mayeur and H. M. Mourad and A. Hunter and M. A. Kenamond",
title = "Coupling continuum dislocation transport with crystal plasticity for application to shock loading conditions",
journal = "Int. J. Plast.",
volume = "76",
pages = "111-129",
year = "2016",
doi = "10.1016/j.ijplas.2015.07.007",
}

@article{Markenscoff:1984JE,
	Author = {Markenscoff, Xanthippi and Ni, Luqun},
	Doi = {10.1007/BF00041084},
	Journal = {J. Elast.},
	Number = {1},
	Pages = {93-95},
	Title = {The transient motion of a screw dislocation in an anisotropic medium},
	Volume = {14},
	Year = {1984},
}

@article{Markenscoff:1984,
author="Xanthippi Markenscoff and Lu Qun Ni",
title="Nonuniform motion of an edge dislocation in an anisotropic solid. {I}",
journal="Quart. Appl. Math.",
year="1984",
volume="41",
pages="475-494",
doi="10.1090/qam/724058",
}

@article{Markenscoff:2001,
author = "Xanthippi Markenscoff and Luqun Ni",
title = "The transient motion of a dislocation with a ramp-like core",
journal = "J. Mech. Phys. Solids",
volume = "49",
number = "7",
pages = "1603-1619",
year = "2001",
doi = "10.1016/S0022-5096(00)00062-4",
}

@article{Pellegrini:2018,
title={Uniformly-moving non-singular dislocations with ellipsoidal core shape in anisotropic media},
author={Pellegrini, Yves-Patrick},
journal = {J. Micromech. Molec. Phys.},
volume = {3},
number = {3 \& 4},
pages = {1840004},
doi = {10.1142/S2424913018400040},
eprint={1808.10272},
archivePrefix={arXiv},
primaryClass={physics.class-ph},
year={2018},
}

@article{Teutonico:1961,
	title = {Dynamical Behavior of Dislocations in Anisotropic Media},
	author = {Teutonico, L. J.},
	journal = {Phys. Rev.},
	volume = {124},
	issue = {4},
	pages = {1039-1045},
	year = {1961},
	doi = {10.1103/PhysRev.124.1039},
}

@article{Ye:2023a,
	title = {Unified crystal plasticity model for fcc metals: {From} quasistatic to shock loading},
	author = {Ye, Changqing and Liu, Guisen and Chen, Kaiguo and Liu, Jingnan and Hu, Jianbo and Yu, Yuying and Mao, Yong and Shen, Yao},
	journal = {Phys. Rev.},
	volume = {B107},
	issue = {2},
	pages = {024105},
	numpages = {11},
	year = {2023},
	doi = {10.1103/PhysRevB.107.024105},
}

@article{Zuanetti:2021,
	author = "Bryan Zuanetti and Darby J. Luscher and Kyle Ramos and Cynthia Bolme and Vikas Prakash",
	title = "{Dynamic flow stress of pure polycrystalline aluminum: Pressure-shear plate impact experiments and extension of dislocation-based modeling to large strains}",
	journal = "J. Mech. Phys. Solids",
	volume = "146",
	pages = "104185",
	year = "2021",
	doi = "10.1016/j.jmps.2020.104185",
}

@book{CRCHandbook,
editor = "John R. Rumble",
title = "CRC Handbook of Chemistry and Physics",
edition = "102nd",
year = "2021",
publisher = "CRC Press",
url = "https://hbcp.chemnetbase.com",
}

@book{Hirth:1982,
  author={John Price Hirth and Jens Lothe},
  title={Theory of Dislocations},
  year={1982},
  edition={second},
  publisher={Wiley},
  address={New York},
}

@article{Katagiri:2023,
	title={Transonic Dislocation Propagation in Diamond},
	author={Katagiri, Kento and Pikuz, Tatiana and Fang, Lichao and Albertazzi, Bruno and Egashira, Shunsuke and Inubushi, Yuichi and Kamimura, Genki and Kodama, Ryosuke and Koenig, Michel and Kozioziemski, Bernard and others},
	eprint={2303.04370},
	archivePrefix = {arXiv},
	primaryClass  = {cond-mat.mtrl-sci},
	journal = {Science},
	volume = {382},
	number = {6666},
	pages = {69-72},
	year = {2023},
	doi = {10.1126/science.adh5563},
}

@article {Gumbsch:1999,
	author = {Gumbsch, Peter and Gao, Huajian},
	title = {Dislocations Faster than the Speed of Sound},
	volume = {283},
	number = {5404},
	pages = {965-968},
	year = {1999},
	doi = {10.1126/science.283.5404.965},
	journal = {Science},
}

@article{Manukhina:2024,
	author = {K. D. Manukhina and V. S. Krasnikov and D. S. Voronin and A. E. Mayer},
	title = {Dislocation activity in aluminum at ultra-high strain rates: {Atomistic} investigation and continuum modeling},
	journal = {Comput. Mater. Sci.},
	volume = {244},
	pages = {113269},
	year = {2024},
	issn = {0927-0256},
	doi = {10.1016/j.commatsci.2024.113269},
}

@article{Olmsted:2005,
  author={Olmsted, David L. and Hector Jr., Louis G. and Curtin, W. A. and Clifton, R. J.},
  title={Atomistic simulations of dislocation mobility in {Al, Ni} and {Al/Mg} alloys},
  journal={Mod. Simul. Mater. Sci. Eng.},
  volume={13},
  pages={371},
  year={2005},
  doi={10.1088/0965-0393/13/3/007},
  archivePrefix = {arXiv},
  eprint = {cond-mat/0412324},
}

@article{Oren:2017,
  author={E. Oren and E. Yahel and G. Makov},
  title={Dislocation kinematics: a molecular dynamics study in {Cu}},
  journal={Mod. Simul. Mater. Sci. Eng.},
  volume={25},
  pages={025002},
  doi={10.1088/1361-651X/aa52a7},
  year={2017},
}

@article{Blaschke:2017lten,
      author         = "Blaschke, Daniel N. and Szajewski, Benjamin A.",
      title          = "Line tension of a dislocation moving through an anisotropic crystal",
      year           = "2018",
      reportNumber   = "LA-UR-17-29936",
      eprint         = "1711.10555",
      archivePrefix  = "arXiv",
      primaryClass   = "cond-mat.mtrl-sci",
      journal        = "Phil. Mag.",
      volume         = "98",
      pages          = "2397-2424",
      doi            = "10.1080/14786435.2018.1489152",
}

@article{Blaschke:2018anis,
      author         = "Blaschke, Daniel N.",
      title          = "Velocity dependent dislocation drag from phonon wind and crystal geometry",
      year           = "2019",
      reportNumber   = "LA-UR-18-22746",
      eprint         = "1804.01586",
      archivePrefix  = "arXiv",
      primaryClass   = "cond-mat.mtrl-sci",
      doi            = "10.1016/j.jpcs.2018.08.032",
      journal        = "J. Phys. Chem. Solids",
      volume         = "124",
      pages          = "24-35",
}

@article{Blaschke:2020MD,
      author      = "Daniel N. Blaschke and Jie Chen and Saryu Fensin and Benjamin Szajewski",
      title          = "Clarifying the definition of `transonic' screw dislocations",
      year          = "2021",
      eprint        = "2008.13760",
      archivePrefix = "arXiv",
      primaryClass  = "cond-mat.mtrl-sci",
      reportNumber  = "LA-UR-20-25514",
      journal = "Phil. Mag.",
      volume  = "101",
      issue = "8",
      pages = "997-1018",
      doi = "10.1080/14786435.2021.1876269",
}

@article{Blaschke:2021impact,
	author      = {Daniel N. Blaschke and Darby J. Luscher},
	title          = {Dislocation drag and its influence on elastic precursor decay},
	year          = {2021},
	reportNumber  = {LA-UR-20-25546},
	eprint={2101.10497},
	archivePrefix={arXiv},
	primaryClass={cond-mat.mtrl-sci},
	journal={Int. J. Plast.},
	volume={144},
	pages={103030},
	doi={10.1016/j.ijplas.2021.103030},
}

@article{Blaschke:2021vcrit,
title={How to determine limiting velocities of dislocations in anisotropic crystals}, 
author={Daniel N. Blaschke},
year={2021},
eprint={2107.01220},
archivePrefix={arXiv},
primaryClass={cond-mat.mtrl-sci},
reportNumber  = {LA-UR-21-26126},
journal={J. Phys.: Cond. Mat.},
volume={33},
pages={503005},
doi={10.1088/1361-648X/ac2970},
}

@article{Dang:2022Mg,
	title={Limiting velocities and transonic dislocations in {Mg}}, 
	author={Khanh Dang and Daniel N. Blaschke and Saryu Fensin and Darby J. Luscher},
	year={2022},
	eprint={2205.11687},
	archivePrefix={arXiv},
	primaryClass={cond-mat.mtrl-sci},
	reportNumber={LA-UR-22-22557},
	journal={Comput. Mater. Sci.},
	volume={215},
	pages={111786},
	doi={10.1016/j.commatsci.2022.111786},
}

@article{Blaschke:2025joss,
	author      = {Daniel N. Blaschke},
	title          = {{PyDislocDyn: A Python} code for calculating dislocation drag and other crystal properties},
	year          = {2025},
	journal       = {JOSS},
	volume        = {10},
	pages         = {9309},
	doi           = {10.21105/joss.09309},
	eprint        = {2509.02900},
	archivePrefix = {arXiv},
	primaryClass  = {cond-mat.mtrl-sci},
	reportNumber  = {LA-UR-25-28917},
}

@online{pydislocdyn,
  author = {Blaschke, Daniel N.},
  title = {{PyDislocDyn}},
  url = {https://github.com/dblaschke-LANL/PyDislocDyn},
  doi = {10.11578/dc.20180619.15},
  OPTversion = {1.3.5},
  year = {2018--2026},
  reportNumber = {C18073},
}

@ARTICLE{Jones:2025MSMSE,
	author = {{Jones}, Kathryn R. and {Dang}, Khanh and {Blaschke}, Daniel N. and {Fensin}, Saryu J. and {Hunter}, Abigail},
	title = "{Exploring the relation between transonic dislocation glide and stacking fault width in {FCC} metals}",
	journal = {Mod. Simul. Mater. Sci. Eng.},
	year = {2025},
	volume = {33},
	number = {2},
	pages = {025020},
	doi = {10.1088/1361-651X/adb017},
	archivePrefix = {arXiv},
	eprint = {2409.10705},
	primaryClass = {cond-mat.mtrl-sci},
}

@article{Akhondzadeh:2023,
	title = {Direct comparison between experiments and dislocation dynamics simulations of high rate deformation of single crystal copper},
	journal = {Acta Materialia},
	volume = {250},
	pages = {118851},
	year = {2023},
	issn = {1359-6454},
	doi = {10.1016/j.actamat.2023.118851},
	author = {Sh. Akhondzadeh and Minju Kang and Ryan B. Sills and K.T. Ramesh and Wei Cai},
}

@article{Tak:2023,
	title = {A discrete dislocation dynamics framework for modeling polycrystal plasticity with hardening},
	journal = {Int. J. Solids Struct.},
	volume = {281},
	pages = {112442},
	year = {2023},
	issn = {0020-7683},
	doi = {10.1016/j.ijsolstr.2023.112442},
	author = {Tawqeer Nasir Tak and Aditya Prakash and Indradev Samajdar and Ahmed Amine Benzerga and P.J. Guruprasad},
}

@article{Bertin:2024,
	author = {Bertin, Nicolas and Bulatov, Vasily V. and Zhou, Fei},
	doi = {10.1038/s41524-024-01378-4},
	isbn = {2057-3960},
	journal = {npj Comp. Mater.},
	number = {1},
	pages = {192},
	title = {Learning dislocation dynamics mobility laws from large-scale {MD} simulations},
	volume = {10},
	year = {2024},
}

@article{Zelenika:2025,
	author = {Zelenika, Albert and Cretton, Adam Andr{\'e}William and Frankus, Felix and Borgi, Sina and Grumsen, Flemming B. and Yildirim, Can and Detlefs, Carsten and Winther, Grethe and Poulsen, Henning Friis},
	doi = {10.1038/s41598-025-88262-3},
	isbn = {2045-2322},
	journal = {Sci. Rep.},
	number = {1},
	pages = {8655},
	title = {Observing formation and evolution of dislocation cells during plastic deformation},
	volume = {15},
	year = {2025},
}

@article{Borgi:2025,
	author = "Borgi, Sina and Winther, Grethe and Poulsen, Henning Friis",
	title = "{Individual dislocation identification in dark-field X-ray microscopy}",
	journal = "J. Appl. Crystallogr.",
	year = "2025",
	volume = "58",
	number = "3",
	pages = "813--821",
	doi = {10.1107/S1600576725002614},
}

@article{Jones1993,
	author = {Jones, D. R. and Perttunen, C. D. and Stuckman, B. E.},
	doi = {10.1007/BF00941892},
	isbn = {1573-2878},
	journal = {J. Optim. Theory Appl.},
	number = {1},
	pages = {157--181},
	title = {Lipschitzian optimization without the {Lipschitz} constant},
	volume = {79},
	year = {1993},
}

\end{document}